\documentclass[smallextended]{svjour3}

\usepackage{graphicx}
\usepackage{dcolumn}
\usepackage{amssymb}
\usepackage{bm}
\usepackage{amsmath}
\usepackage{subfigure}
\usepackage{longtable}
\makeatother

\begin{document}

\title{Bianchi $VII_A$ solutions of effective quadratic gravity}

\author{Juliano A. de Deus \and Daniel M\"uller}
\institute{J. A. de Deus \at Instituto de F\'\i sica, UnB, 
Campus Universit\'ario Darcy Ribeiro, 
Cxp 04455, 70919-970, Bras\'\i lia DF, Brazil\\
\email{julianoalves@fis.unb.br}\and
D. M\"uller \at 
Instituto de F\'\i sica, UnB, 
Campus Universit\'ario Darcy Ribeiro,
Cxp 04455, 70919-970, Bras\'\i lia DF, Brazil\\
\email{muller@fis.unb.br}
} 
\maketitle
\begin{abstract}
It is believed that soon after the Planck time, Einstein's general relativity theory should be corrected to an effective quadratic theory. Numerical solutions for the anisotropic generalization of the Friedmann ``open" model $H^ 3$ for this effective gravity are given. It must be emphasized that although numeric, these solutions are exact in the sense that they depend only on the precision of the machine. 
The solutions are identified asymptotically in a certain way. It is found solutions which asymptote de Sitter space, Riemann flat space and a singularity. The question of isotropisation of an initially anisotropic Universe is of great importance in the context of cosmology. Although isotropisation is not directly discussed in this present work, we show that sufficiently small anisotropies, do not increase indefinitely according to particular quadratic gravity theories. It can be understood as weak isotropisation, and we stress that this result is strongly dependent on initial conditions. 
\end{abstract}
\section{Introduction}
\newcommand{\imsize}{0.83\columnwidth} 
\newcommand{\halfsize}{0.38\columnwidth} 
The semi-classical theory considers the back reaction of quantum fields in a classical geometric 
background. It began about forty years ago with De Witt \cite{DeWitt}, and since then, its 
consequences and applications are still under research, see for example \cite{Hu}.

Different from the usual Einstein-Hilbert action, the one loop effective gravitational action surmounts 
to quadratic theories in curvature, see for example \cite{DeWitt}, \cite{liv}. It is the gravitational version of the Heisenberg-Euler electromagnetism. As it is well known, vacuum polarization introduces non linear corrections into Maxwell electrodynamics \cite{schwinger}, first obtained by Heisenberg- Euler \cite{heisenberg}.

This quadratic gravity was previously studied by Starobinsky \cite{S}, and more recently, also 
by others \cite{Shapiro}. The higher order terms are in fact responsible for a slowly decaying dark energy. In \cite{Shapiro} only the homogeneous and isotropic space 
time is studied. It is of interest for example in the context of the final stages of evaporation of black holes,  inflationary theories \cite{coule}, in the approach to the singularity \cite{montani} and also in a more theoretical context \cite{cotsakis}. 

The effective gravity was apparently first
investigated in Tomita's article \cite{berkin} for general Bianchi $I$
spaces. They found that the presence of anisotropy contributes to the formation of the singularity. Gurovich and Starobinsky studied the fourth-order $f(R)$, in which the anisotropic part of the metric was integrated explicitly. Berkin's work shows that a quadratic
Weyl theory is less stable than a quadratic Riemann scalar $R^2$. In particular, in the very interesting article \cite{barrow-hervik}, Barrow and Hervik addressed the anisotropic cases of Bianchi $I$ and 
$II$. The most interesting results of Barrow and Hervik are the exact solutions for quadratic theories of the same type investigated in this present article. 
Instead of the metric, the field equations in \cite{barrow-hervik} are written in a different set of variables which by now is a well known procedure used in the context of dynamical systems in cosmology \cite{dsc}. We intend to switch to these set of variables in future works. Homogenous solutions in the context of quadratic gravity was also addressed by \cite{resto}. Schmidt did a review of higher order gravity theories in connection to cosmology \cite{hjs}. 

Also in the context of quadratic theories we have the interesting alternative of Gauss-Bonnet type $F(G)$ by Odintsov, Nojiri and collaborators \cite{odintsov06}, for recent reviews see \cite{odintsov}. When the action depends on a arbitrary function of the Gauss-Bonnet term, it is not a top invariant and a consistent dynamic follows from it. Theories of $R^2$ type are also investigated, for example in \cite{shapiro-odintsov}. 

In our case the Gauss-Bonnet term is understood as a surface term and discarded. In fact, using the Gauss-Bonnet term and the relation between the Weyl tensor and Riemann tensor it is possible to demonstrate the equivalence between the particular theory under investigation in this work and other quadratic theories of gravity. 

We have previously numerically investigated Bianchi $I$ type solutions  \cite{sandro}. Soon after this 
result we also investigated the stability of the this particular Bianchi $I$ case \cite{daniel-sandro}. It 
turns out that for zero cosmological constant $\Lambda=0$, Minkowski geometry is structurally stable in 
the sense that there is a basin of attraction to Minkowski solution.  And for $\Lambda >0 $ de Sitter type 
geometry is structurally stable, also in the same sense. Thus, soon after the Planck era, the effective 
one loop quadratic gravity predicts for the very particular Bianchi I models, that there is a basin of 
attraction to Minkowski space for $\Lambda=0$ and a basin of attraction to de Sitter space for $\Lambda>0$, see \cite{daniel-sandro}. 

Within Einstein's general relativity theory, Wainwright et al \cite{BVII_Isotropy} discovered  
that almost any initial condition for Bianchi $VII_A$ evolve to an exact plane wave solution. For a very good and recent review on homogenous models in the context of Einstein's theory, see Ellis \cite{BiancRevisEllis}.

In this present work, the solutions for the Bianchi $VII_A$ spatial homogenous space for the effective quadratic gravity is addressed. This is the anisotropic generalization 
of the hyperbolic $H^3$ Friedmann ``open" space. For any spatially homogenous space-time, the 
dynamical equations of motion, of any theory, result in a non linear system of ordinary differential equations. 
For the particular quadratic gravity, the ordinary differential equation is of degree $4$. 

The initial conditions chosen are very near to exact solutions of the quadratic theory, and also only classical vacuum source is considered, which we believe is the best description soon after the Planck era. The solutions are obtained numerically and they are understood asymptotically in a certain sense described in the text. It is shown that depending on the initial conditions and parameters, the de Sitter solution, or Riemann flat solution, or a singularity type solution is obtained asymptotically. 

The paper is organized as follows. In section 2 a brief review of the coordinate free method is used to obtain the field equations for the Bianchi $VII_A$ case, and also the Newman-Penrose coefficients. In Section 3 $3$ different initial conditions are analyzed: one that converges to de Sitter space, one that converges to Riemann flat space, and one that converges to a singularity. In the last section we present our conclusions. We believe the index notation should be clear from the text. 

\section{Theory and Development}

For completeness, we give a brief overview about the tetrad basis, coordinate free description
of spatially homogenous geometries, for further details see for example \cite{ExactSolutionsStephani}. 

The right-invariant basis which describes the group of isometries
$G_{3}$ are the Killing vectors that follow from the relation%
\begin{equation}
\left[\xi_{a},\xi_{b}\right]=C_{\: ab}^{c}\xi_{c},\label{eq commutation of killing vector}\end{equation}
where $C_{\:[ab]}^{c}$ are the structure constants of the group which
satisfies the Jacobi identity\begin{equation}
C_{\:[ab}^{d}C_{\: c]d}^{f}=0.\label{eq jacobi identity for structure constant}\end{equation}
Since right and left actions commute,\begin{equation}
\left[\eta_{a},\xi_{b}\right]=0,\label{eq null commutation between eta-csi}\end{equation}
where $\{\eta_{a}\}$ is the left-invariant vector basis. The right-invariant
1-form basis $\{\omega^{a}\}$, its dual, follows from the first Cartan
equation\begin{equation}
d\omega^{a}=-\Gamma_{\: b}^{a}\wedge\omega^{b}=\Gamma_{\: bc}^{a}\omega^{b}\wedge\omega^{c},\label{eq first Cartan}\end{equation}
where the connection 1-form is defined by $\Gamma_{\: b}^{a}=\Gamma_{\: bc}^{a}\omega^{c}$.
Since $2\Gamma_{\:[bc]}^{a}=C_{\: bc}^{a}$, then\begin{equation}
d\omega^{a}=\frac{1}{2}C_{\: bc}^{a}\omega^{b}\wedge\omega^{c}.\label{eq diferential equation for one-form basis}\end{equation}

The Bianchi classification of all possible three-dimension algebras, based on the structure constants,
can be obtained by Sch\"ucking's method. Once the structure constants
are known, the vector and 1-form basis are obtained in the following
way: the first-order differential equation (\ref{eq diferential equation for one-form basis})
is solved to determine the basis $\{\omega^{a}\}$ and the duality
relation $\omega^{a}\eta_{b}=\delta_{\: b}^{a}$ gives the basis $\{\eta_{a}\}$.
Since $\mathcal{L}_{\xi_{a}}(\omega^{b}\eta_{c})=(\mathcal{L}_{\xi_{a}}\omega^{b})\eta_{c}+\omega^{b}(\mathcal{L}_{\xi_{a}}\eta_{c})=\mathcal{L}_{\xi_{a}}(\delta_{\: c}^{b})=0$
and $\mathcal{L}_{\xi_{a}}\eta_{b}=[\xi_{a},\eta_{b}]=0$ (from (\ref{eq null commutation between eta-csi})),
then the Killing vectors $\{\xi_{a}\}$ are obtained by solving the following
first-order differential equation%
\begin{equation}
\mathcal{L}_{\xi_{a}}\omega^{b}=(\omega_{\: i,j}^{b}\xi_{a}^{\: j}+\omega_{\: j}^{b}\xi_{a,\: i}^{\: j})dx^{i}=0.\end{equation}
The structure constants and the basis for the type Bianchi $VII_A$
($A$ is the group parameter) are shown Table \ref{tab BVIIA basis and structure constants}.

\begin{table}
\caption{$BVII_A$ basis and structure constants.\label{tab BVIIA basis and structure constants}
}
\begin{tabular}{cccc}
\hline
$\xi_{1}$ & $\partial_{x}+(z-Ay)\partial_{y}-(y+Az)\partial_{z}$ & $C_{\:12}^{2}$ & $A$\\
$\xi_{2}$ & $\partial_{y}$ & $C_{\:21}^{2}$ & $-A$\\
$\xi_{3}$ & $\partial_{z}$ & $C_{\:13}^{3}$ & $A$\\
\hline
$\eta_{1}$ & $\partial_{x}$ & $C_{\:31}^{3}$ & $-A$\\
$\eta_{2}$ & $e^{-Ax}(\cos(x)\partial_{y}-\sin(x)\partial_{z})$ & $C_{\:13}^{2}$ & $-1$\\
$\eta_{3}$ & $e^{-Ax}(\sin(x)\partial_{y}+\cos(x)\partial_{z})$ & $C_{\:31}^{2}$ & $1$\\
\hline
$\omega^{1}$ & $dx$ & $C_{\:12}^{3}$ & $1$\\
$\omega^{2}$ & $e^{Ax}(\cos(x)dy-\sin(x)dz)$ & $C_{\:21}^{3}$ & $-1$\\
$\omega^{3}$ & $e^{Ax}(\sin(x)dy+\cos(x)dz)$ & $C_{\: bc}^{a}$ (else) & $0$\\
\hline
\end{tabular}
\end{table}
The 1-form basis performs the correct projection to obtain a spatially
homogeneous four-space%
\begin{equation}
ds^{2}=g_{ab}(t)\omega^{a}\otimes\omega^{b}=-dt^{2}+g_{\alpha\beta}(t)\omega^{\alpha}\otimes\omega^{\beta}.\label{eq line element for spatially homogeneous four-space}
\end{equation}
In this new basis, it can be seen that, indeed, $\xi_{a}$ is a Killing
vector:
\begin{equation}
\mathcal{L}_{\xi_{c}}g=\mathcal{L}_{\xi_{c}}(g_{ab}\omega^{a}\otimes\omega^{b})=(\mathcal{L}_{\xi_{c}}g_{ab})\omega^{a}\otimes\omega^{b}+g_{ab}(\mathcal{L}_{\xi_{c}}\omega^{a})\otimes\omega^{b}+g_{ab}\omega^{a}\otimes(\mathcal{L}_{\xi_{c}}\omega^{b})=0,
\end{equation}
remind that $g_{ab}(t)$ is a function of $t$ only. The partial derivatives
$g_{ij,k}$ are replaced by directional derivatives $g_{ab|c}=g_{ab,i}\eta_{c}^{\: i}$,
which are projections of the derivatives in the basis $\{\eta_{a},\omega^{a}\}$.
The covariant derivative written in tetrad indices is\begin{equation}
\nabla_{a}T^{b}=T_{\:|a}^{b}+\Gamma_{\: ca}^{b}T^{c},\label{eq covariant derivative}\end{equation}
where the torsion-free connection follows from $\nabla_{a}(g_{bc}\omega^{b}\otimes\omega^{c})=0$
and (\ref{eq first Cartan}),\begin{equation}
\Gamma_{abc}=\frac{1}{2}(g_{ab|c}+g_{ac|b}-g_{bc|a}+C_{abc}-C_{cab}-C_{bac}).\label{eq conexion in terms of g and C}\end{equation}
The Riemann tensor is obtained from\begin{equation}
[\nabla_{a},\nabla_{b}]T^{c}=R_{\: dab}^{c}T^{d},\end{equation}
which according to (\ref{eq covariant derivative}) and (\ref{eq conexion in terms of g and C})
results in
\begin{equation}
R_{\: bcd}^{a}=\Gamma_{\: bd|c}^{a}-\Gamma_{\: bc|d}^{a}+\Gamma_{\: fc}^{a}\Gamma_{\: bd}^{f}-\Gamma_{\: fd}^{a}\Gamma_{\: bc}^{f}+C_{\: cd}^{f}\Gamma_{\: bf}^{a}.\label{eq Riemann tensor with C}.
\end{equation}
The Ricci tensor and Ricci scalar are obtained as usual: $R_{ab}=R_{\: acb}^{c}$
and $R=R_{\: a}^{a}$.

The field equations for the semiclassical theory are obtained performing
metric variations in the gravitational Lagrangian \cite{sandro}, \cite{daniel-sandro}
\begin{equation}
\mathcal{L}=\sqrt{-g}\left[-\Lambda+R+\alpha\left(R_{ab}R^{ab}-\frac{1}{3}R^{2}\right)+\beta R^{2}\right]+\mathcal{L}_{q},
\label{acao}
\end{equation}
where $\mathcal{L}_{q}$ is the quantum part of Lagrangian and $\alpha$
and $\beta$ are constants. For the spatially homogenous space they are described
by the tensor $E=E_{ab}\omega^{a}\otimes\omega^{b}$, $\omega^4=dt$,
\begin{equation}
E_{ab}\equiv G_{ab}+\frac{1}{2}g_{ab}\Lambda-\left(\beta-\frac{1}{3}\alpha\right)H_{\: ab}^{(1)}-\alpha H_{\: ab}^{(2)}-T_{ab}=0,\label{eq field equation for semiclassical theory}
\end{equation}
where
\begin{eqnarray*}
&&G_{ab}=R_{ab}-\frac{1}{2}g_{ab}R,\\
&&H_{ab}^{(1)}=\frac{1}{2}g_{ab}R^{2}-2RR_{ab}-2g_{ab}\square R+2R_{;ab},\\
&&H_{ab}^{(2)}=\frac{1}{2}g_{ab}R^{cd}R_{cd}-\square R_{ab}-\frac{1}{2}g_{ab}\square R+R_{;ab}-2R^{cd}R_{cbda}
\end{eqnarray*}
and the connection used is \eqref{eq conexion in terms of g and C}, the Riemann tensor is \eqref{eq Riemann tensor with C} and $\square=\nabla_{a}\nabla^{a}$ is D'Alembertian operator. 

The second quantization program of quadratic gravity is provided by Stelle \cite{Stelle}. According to \cite{Stelle}, the linearized filed equations result in eight degrees of freedom. Two correspond to the familiar massless spin 2 graviton. Five correspond to a massive spin 2 particle with mass $m_2=1/{\sqrt{\alpha}}$. The last degree of freedom correspond to a massive scalar particle  with mass $m_0=1/{\sqrt{-6\beta}}$. In this work we choose $\alpha>0$ and $\beta<0$ throughout. The presence of tachyons occur if $\alpha<0$ or $\beta>0$ and in our case, would indicate the linear instability of the solution. The existence of oscillations in our solutions is consistent with a normal scalar and a tensor massive particles. 

The counterterms $R^2$, $R_{ab}R^{ab}$, $\Lambda$  and $R$ in \eqref{acao} are precisely the ones necessary in order to obtain a finite vacuum expectation  value of the energy momentum tensor, see for example \cite{christensen}. A theory without these counterterms is inconsistent from the point of view of the renormalization of the quantum field in $\mathcal{L}_{q}$. The renormalized vacuum expectation value of the energy momentum tensor is set to zero, which emphasizes the effects of a theory that should have been considered from the start. We are disregarding any classical source in this work. 

The finite contributions to the vacuum expectation value of the energy momentum tensor are known only for very particular situations, split rank spaces, for which the heat kernel can be obtained exactly \cite{camporesi}, then the point splitting method gives $\langle 0|T_{ab}| 0 \rangle$ exactly. These solutions are known as self consistent, $G_{ab}=\langle 0|T_{ab}| 0 \rangle$ and there are very few cases, for example \cite{dowker}. 

As in any other metric theory, the covariant divergence of $E_{ab}$
must be zero
\[
\nabla^{a}E_{ab}=0.
\]
From $\nabla^{a}E_{ab}=0$ it follows that if $E_{44}=0$ and $E_{4\alpha}=0$
initially, then they will remain zero fo any time, and act as constraints
on the initial conditions. Consequently, these constraints are checked
to test the accuracy of the numerical results, while $E_{\alpha\beta}=0$
represent the real dynamical equations of the problem (see for example
\cite{Stephani}, p. 165).

The homogeneous and anisotropic metric supposed throughout this work in the numerical results is the following
way,\begin{equation}
g_{ab}(t)=\left(\begin{array}{cccc}
a_{1}^{\:2}(t)&0 & 0 & 0\\
0 & a_{2}^{\:2}(t) & a_{4}(t)&0\\
0 & a_{4}(t) & a_{3}^{\:2}(t)&0\\
0&0 &0&-1
\end{array}\right),\label{eq homogeneous and anisotropic line element}\end{equation}
\begin{equation}
ds^{2}=-dt^{2}+g_{\alpha\beta}(t)\omega^{\alpha}\otimes\omega^{\beta},\label{eq line element for spatially homogeneous four-space again}\end{equation}
where $g_{\alpha\beta}$ is the spatial part, and the Bianchi $VII_A$ 1-form basis is given in Table
\ref{tab BVIIA basis and structure constants}. The replacement
of the above line element in (\ref{eq field equation for semiclassical theory})
taking into account the appropriate connection and covariant derivative
defined in (\ref{eq conexion in terms of g and C}) and (\ref{eq covariant derivative})
results in a non-linear fourth-order ordinary differential equation
system in the functions $a_{i}(t)$, $i=1,2,3,4$:\begin{equation}
\frac{d^{4}}{dt^{4}}a_{i}(t)=f_{i}\left(\dddot{a}_{j}(t),\ddot{a}_{j}(t),\dot{a}_{j}(t),a_{j}(t)\right).\end{equation}

It is well known that the Bianchi $VII_A$ type
contains the hyperbolic isotropic model. Replacing the Bianchi $VII_A$
basis Table \ref{tab BVIIA basis and structure constants} and instead
of (\ref{eq homogeneous and anisotropic line element}) the isotropic
metric $g_{\alpha\beta}(t)=\mbox{diag}(a^{\:2}(t),a^{\:2}(t),a^{\:2}(t))$
into the line element (\ref{eq line element for spatially homogeneous four-space again})
results in
\begin{equation}
ds^{2}=-dt^{2}+a^{\:2}(t)[dx^{2}+e^{2Ax}(dy^{2}+dz^{2})].
\label{elisotropico}
\end{equation}
With the coordinate transformation
\begin{eqnarray*}
&&x  =  \frac{\ln(\cosh\chi-\sinh\chi\cos\theta)}{A},\\
&&y  =  \frac{\sin\theta\cos\varphi}{A\coth\chi-\cos\theta},\\
&&z  =  \frac{\sin\theta\sin\varphi}{A\coth\chi-\cos\theta}
\end{eqnarray*}
the usual Friedmann line element is obtained,
\begin{equation}
ds^{2}=-dt^{2}+\frac{a^{\:2}(t)}{A^{2}}[d\chi^{2}+\sinh^{2}\chi(d\theta^{2}+\sin^{2}\theta d\varphi^{2})].\label{eq Friedmann hyperbolic line element}
\end{equation}

First let us emphasize that every Einstein space, satisfying $R_{ab}=\Lambda g_{ab}/2$, is an exact solution of this one loop quadratic effective theory given in \eqref{eq field equation for semiclassical theory}. Note the particular case when the constant $\Lambda=0$: vacuum solutions of Einstein's equations are also exact solutions of \eqref{eq field equation for semiclassical theory}. So there is the following exact solution of \eqref{eq field equation for semiclassical theory}
for Bianchi $VII_A$ with the diagonal metric $g_{\alpha\beta}(t)=\mbox{diag}(a_{1}^{\:2}(t),a_{2}^{\:2}(t),a_{3}^{\:2}(t))$
\begin{eqnarray}
&&a_1(t)=A\sqrt{\frac{6}{\Lambda}}\sinh\left(\sqrt{\frac{\Lambda}{6}}(t+C_1/A)\right) \nonumber\\
&&a_2(t)=AC_2\sqrt{\frac{6}{\Lambda}}\sinh\left(\sqrt{\frac{\Lambda}{6}}(t+C_1/A)\right)\nonumber\\
&&a_3(t)=AC_2\sqrt{\frac{6}{\Lambda}}\sinh\left(\sqrt{\frac{\Lambda}{6}}(t+C_1/A)\right), \label{soldeSitter}
\end{eqnarray}
where $C_{1}$ and $C_{2}$ are integration constants and $A$ is
the group parameter. This solution is in fact the de Sitter solution modulo a coordinate transformation. The particular case when $\Lambda\rightarrow 0$,  \cite{DissertJuliano},
\begin{eqnarray*}
&&a_{1}(t)=At+C_{1},\\
&&a_{2}(t)=C_{2}(At+C_{1}),\\
&&a_{3}(t)=C_{2}(At+C_{1}),\\
&&A\neq0,
\label{eq Juliano exact solution}
\end{eqnarray*}
which is equivalent to Minkowski space since the Riemann tensor is identically zero, $R_{\: bcd}^{a}=0$. Morever, it is in fact Milne space 
\[
ds^{2}=-dt^{2}+t^{2}[d\chi^{2}+\sinh^{2}\chi(d\theta^{2}+\sin^{2}\theta d\varphi^{2})],
\]
modulo a coordinate transformation. 

The intention is to characterize the obtained numerical solutions in some way. The Weyl tensor follows from the Riemann tensor (\ref{eq Riemann tensor with C})
as\begin{equation}
C_{abcd}=R_{abcd}-\frac{1}{2}(g_{ca}R_{bd}+g_{db}R_{ca}-g_{cb}R_{da}-g_{da}R_{cb})+\frac{1}{6}R(g_{ca}g_{db}-g_{cb}g_{da}).
\label{weyl-riemann}
\end{equation}

A complex null basis can be defined 
\begin{eqnarray}
&&k^{a}k_{a}=k^{a}t_{a}=k^{a}\bar{t}_{a}=l^{a}l_{a}=l^{a}t_{a}=l^{a}\bar{t}_{a}=t^{a}t_{a}=\bar{t}^{a}\bar{t}_{a}=0,\nonumber\\
&&t^{a}\bar{t}_{a}=-k^{a}l_{a}=1,
\label{basenula}
\end{eqnarray}
with the corresponding null metric 
\begin{equation}
\tilde{g}_{AB}=g_{ab}A^aB^b=\left(\begin{array}{cccc}
0&-1 & 0 & 0\\
-1& 0 & 0&0\\
0 & 0&0&1\\
0&0 &1&0
\end{array}\right),
\label{metricanula}
\end{equation}
where $A^a$ and $B^b$ are the null vectors in \eqref{basenula}. The Newman-Penrose complex coefficients are in fact the tetrad components of the 
Weyl tensor 
\begin{eqnarray*}
&&\psi_{0}=C_{abcd}k^{a}t^{b}k^{c}t^{d},\\
&&\psi_{1}=C_{abcd}k^{a}l^{b}k^{c}t^{d},\\
&&\psi_{2}=C_{abcd}k^{a}t^{b}\bar{t}^{c}l^{d},\\
&&\psi_{3}=C_{abcd}k^{a}l^{b}\bar{t}^{c}l^{d},\\
&&\psi_{4}=C_{abcd}\bar{t}^{a}l^{b}\bar{t}^{c}l^{d}.
\end{eqnarray*}
Also the Ricci tensor can be projected to the null tetrad, resulting into the independent components,
\begin{eqnarray}
&&R_{kk}=R_{ab}k^ak^b\nonumber\\
&&R_{kl}=R_{ab}k^al^b\nonumber\\
&&R_{ll}=R_{ab}l^al^b\nonumber\\
&&R_{kt}=R_{ab}k^at^b\nonumber\\
&&R_{lt}=R_{ab}l^at^b\nonumber\\
&&R_{tt}=R_{ab}t^at^b\nonumber\\
&&R_{t\bar{t}}=R_{ab}t^a\bar{t}^b, \label{riccinull}
\end{eqnarray}
the first $3$ and the last one are real, and the remaining ones are arbitrary complex numbers comprising of course 10 independent components. 

The Newman-Penrose coefficients are related to the Petrov classification as shown for example
in \cite{ExactSolutionsStephani} and \cite{Stephani}. When all the $\psi$'s
are zero it's a Petrov type O, which characterizes a conformal Minkowski  
space and the Weyl tensor vanishes. An exactly conformally flat solution is either a generalized Friedmann solution or an interior Schwarzschild solution as it can be seen for example in \cite{ExactSolutionsStephani} pg. 413. Minkowski  and de Sitter geometries, are particular cases of conformally flat solutions. 

Before getting into the numeric result, we call the attention to the book \cite{dsc}, pgs. 62- 64. As it is well written there, there is not a complete statement as to what constitutes a minimal set for ensuring that a cosmological model is close to Friedmann-Lemaitre model. This book is based on an article \cite{Stoeger1995} for which some assumptions are made: i) Einstein's equations are satisfied. ii) the source is a mixture of radiation and dust. 

We are not concerned if the two above conditions are satisfied in this present work. So we do not expect that their result, albeit being very interesting, should be verified in the particular context being discussed here. 

In the following we will present a particular example. Consider the Bianchi $VII_A$ metric written in the appropriate base given in Table \ref{tab BVIIA basis and structure constants}
\begin{equation}
g_{ab}=\left(\begin{array}{cccc}
a(t)^2&0 & 0 & 0\\
0& b(t)^2 & 0&0\\
0 & 0&b(t)^2 &0\\
0&0 &0&-1
\end{array}\right).  
\label{cexemplo}
\end{equation}
The time like vector $u^a=(0,0,0,1)$ is geodesic and orthogonal to the group orbit. The magnetic part $H_{ab}=0$, and the electric $E_{ab}$ part of the Weyl tensor is
\[ 
E_a^b=\left(\begin{array}{cccc}
2\psi_2&0 & 0 & 0\\
0& -\psi_2 & 0&0\\
0 & 0&-\psi_2 &0\\
0&0 &0&0
\end{array}\right),
\]
where $\psi_2$ is the above defined Newman-Penrose coefficient 
\begin{equation}
\psi_2=\frac{1}{6}\left( \frac{-b(t)^2\ddot{a}(t)+b(t)\dot{b}(t)\dot{a}(t)+b(t)a(t)\ddot{b}(t)-a(t)\dot{b}(t)^2
}{a(t)b(t)^2}\right).
\label{psi2}
\end{equation}
Also regarding \eqref{cexemplo}, the expansion $H=1/3\Theta$, $\Theta=\nabla_c u^c$ is
\begin{equation}
H=\frac{1}{3}\frac{\dot{a}(t)}{a(t)}+\frac{2}{3}\frac{\dot{b}(t)}{b(t)}.
\label{expansao}
\end{equation}
Now consider 
\begin{eqnarray}
a(t)=\frac{tA}{1-t^{-1/3}\sin (t)} \nonumber\\
b(t)=\frac{tA}{1-t^{-1/4}\cos (t)}.
\label{fcexemplo}
\end{eqnarray}
>From \eqref{fcexemplo} it is easily seen that \eqref{cexemplo} as $t\rightarrow \infty$ is
\begin{equation}
\lim_{t\rightarrow \infty}g_{ab}=\left(\begin{array}{cccc}
t^2A^2&0 & 0 & 0\\
0& t^2A^2 & 0&0\\
0 & 0&t^2A^2 &0\\
0&0 &0&-1
\end{array}\right),
\label{cexassint}
\end{equation}
which according to \eqref{eq Friedmann hyperbolic line element} and \eqref{elisotropico} is Milne's space which is identical to the Minkowski solution. 

Considering \eqref{fcexemplo} and \eqref{cexemplo}, now we can explicitly obtain the asymptotic $t\rightarrow \infty$ limits of the Newman-Penrose coefficient \eqref{psi2} and the expansion \eqref{expansao} to be 
\begin{eqnarray}
&&\lim_{t\rightarrow \infty}\psi_2=-\frac{1}{6}t^{-1/4}\cos (t)\nonumber\\
&&\lim_{t\rightarrow \infty}E_{ab}E^{ab}=\frac{1}{6}t^{-1/2}\cos (t)^2\nonumber\\
&&\lim_{t\rightarrow \infty} H=\frac{2}{3}t^{-1/4}\sin (t)
\label{psi,H,assintoticos}
\end{eqnarray}
 
This example shows that  Definition 2.1 of \cite{dsc} 
\begin{eqnarray}
\sqrt{H_{ab}H^{ab}}/H^2<<\epsilon\nonumber\\
\sqrt{E_{ab}E^{ab}}/H^2 << \epsilon 
\label{cellis}
\end{eqnarray}
$0<\epsilon<<1$ is not necessary for the asymptotic approach to a Friedmann-Lemaitre model. Of course the authors of \cite{dsc} were aware of this fact as it can be read on pg. 64 where they carefully write that: if Definition 2.1 is satisfied, then the metric can be locally written in an almost-RW form.

As a final word we will use a different criteria, namely, inspired in \eqref{psi,H,assintoticos}, the asymptotic tetrad components of the Weyl and Ricci tensor
\begin{eqnarray}
&&\lim_{t\rightarrow \infty}C_{abcd}<<\epsilon \nonumber\\
&&\lim_{t\rightarrow \infty}R_{ab}
\label{criterio}
\end{eqnarray}
$0<\epsilon<<1$ which is less restrictive than \eqref{cellis}. We emphasize that the asymptotic behavior of the tetrad components \eqref{criterio} is a weaker criteria than the one given in Definition 2.1 of \cite{dsc}. 

For instance, the Kasner solution \cite{Kasner} and the exact Bianchi $VII_A$ gravity wave \cite{ondaexata} obey \eqref{criterio} with $R_{ab}\equiv 0$: on the other hand these two exact solutions do not satisfy \eqref{cellis}. So in principle our criteria would not distinguish between Kasner exact anisotropic solution, and for example, the limit  as $t\rightarrow \infty$ given in \eqref{cexassint}, with \eqref{fcexemplo} and \eqref{cexemplo}, which is clearly a space for which isotropisation occurs in the strong sense. 

\section{Numerical Solutions}
\subsection{Asymptotically de Sitter Solution}
We choose an initial condition near the exact de Sitter \eqref{soldeSitter} solution with $t=1$ and $C_1=0$, $C_2=20$, $A=0.1$, $\alpha=1$, $\beta=-5.0$ and $\Lambda=0.02$. The only non null coefficients consistent with the 
$E_{44}\equiv 0$ and $E_{41}\equiv 0$ constraints are 
\begin{eqnarray}
&&a_1(t)=A\sqrt{\frac{6}{\Lambda}}\sinh\left(\sqrt{\frac{\Lambda}{6}}t\right) - 2.8\times 10^{-3}\nonumber\\
&&a_2(t)=C_2A\sqrt{\frac{6}{\Lambda}}\sinh\left(\sqrt{\frac{\Lambda}{6}}t\right)\nonumber\\
&&a_3(t)=C_2A\sqrt{\frac{6}{\Lambda}}\sinh\left(\sqrt{\frac{\Lambda}{6}}t\right)\nonumber\\
&&\dot{a_1}(t)=A\cosh\left(\sqrt{\frac{\Lambda}{6}}t\right)\nonumber\\
&&\dot{a_2}(t)=C_2A\cosh\left(\sqrt{\frac{\Lambda}{6}}t\right)\nonumber
\end{eqnarray}
\begin{eqnarray}
&&\dot{a_3}(t)=C_2A\cosh\left(\sqrt{\frac{\Lambda}{6}}t\right)\nonumber\\
&&\ddot{a_1}(t)=A\sqrt{\frac{\Lambda}{6}}\sinh\left(\sqrt{\frac{\Lambda}{6}}t\right)\nonumber\\
&&\ddot{a_2}(t)=C_2A\sqrt{\frac{\Lambda}{6}}\sinh\left(\sqrt{\frac{\Lambda}{6}}t\right)\nonumber\\
&&\ddot{a_3}(t)=C_2A\sqrt{\frac{\Lambda}{6}}\sinh\left(\sqrt{\frac{\Lambda}{6}}t\right)\nonumber\\
&&\dddot{a}_1(t)= -1.14153106\times 10^{-2}\nonumber\\
&&\dddot{a}_2(t)=C_2A\frac{\Lambda}{6}\cosh\left(\sqrt{\frac{\Lambda}{6}}t\right)\nonumber\\
&&\dddot{a}_3(t)=  -4.65026064\times 10^{-1},\label{cndl}
\end{eqnarray}
we stress that numerically the initial condition $a_4=0$ but it rapidly increases. The metric does not remains diagonal. 

\begin{figure}[htpb]
 \begin{center}
 \resizebox{\imsize}{!}{\includegraphics{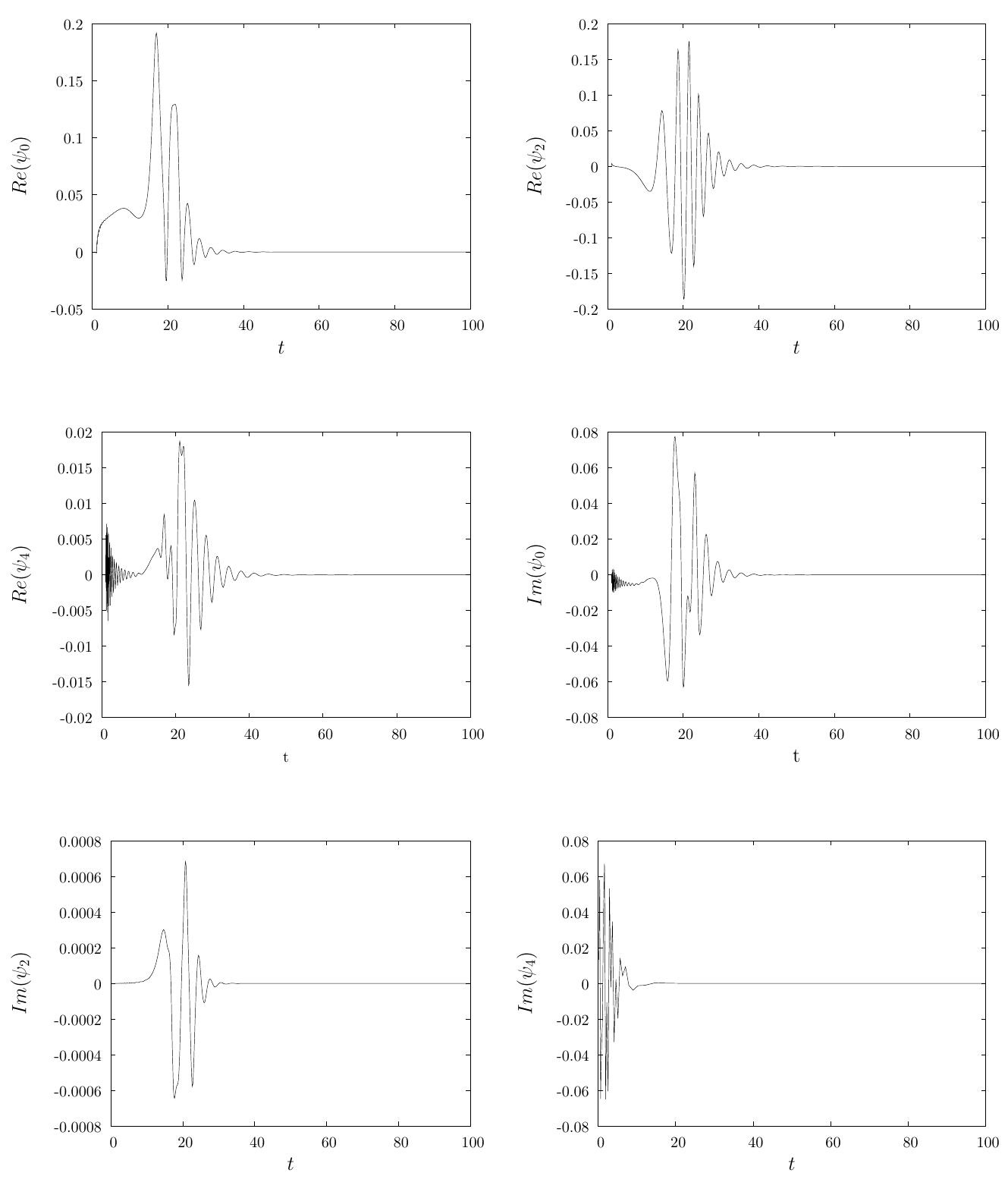}}
  \end{center}
  \caption{Numerical evolution of the real and imaginary parts of the Newman-Penrose coefficients $\psi_0$ - $\psi_4.$ For the initial condition given in the text. The numerical integration was done until the proper time $t=100$.}
  \label{figura1l}
\end{figure}

The time evolution of the non zero Newman-Penrose coefficients, $\psi_0$-$\psi_4$ are shown in Fig.  \ref{figura1l}. For the line element chosen $\psi_1\equiv \psi_3\equiv 0$. It can be seen that asymptotically the Newman-Penrose coefficients all vanish so the Weyl tensor is asymptotically zero. 

\begin{figure}[htpb]
  \begin{center}
   \resizebox{\imsize}{!}{\includegraphics{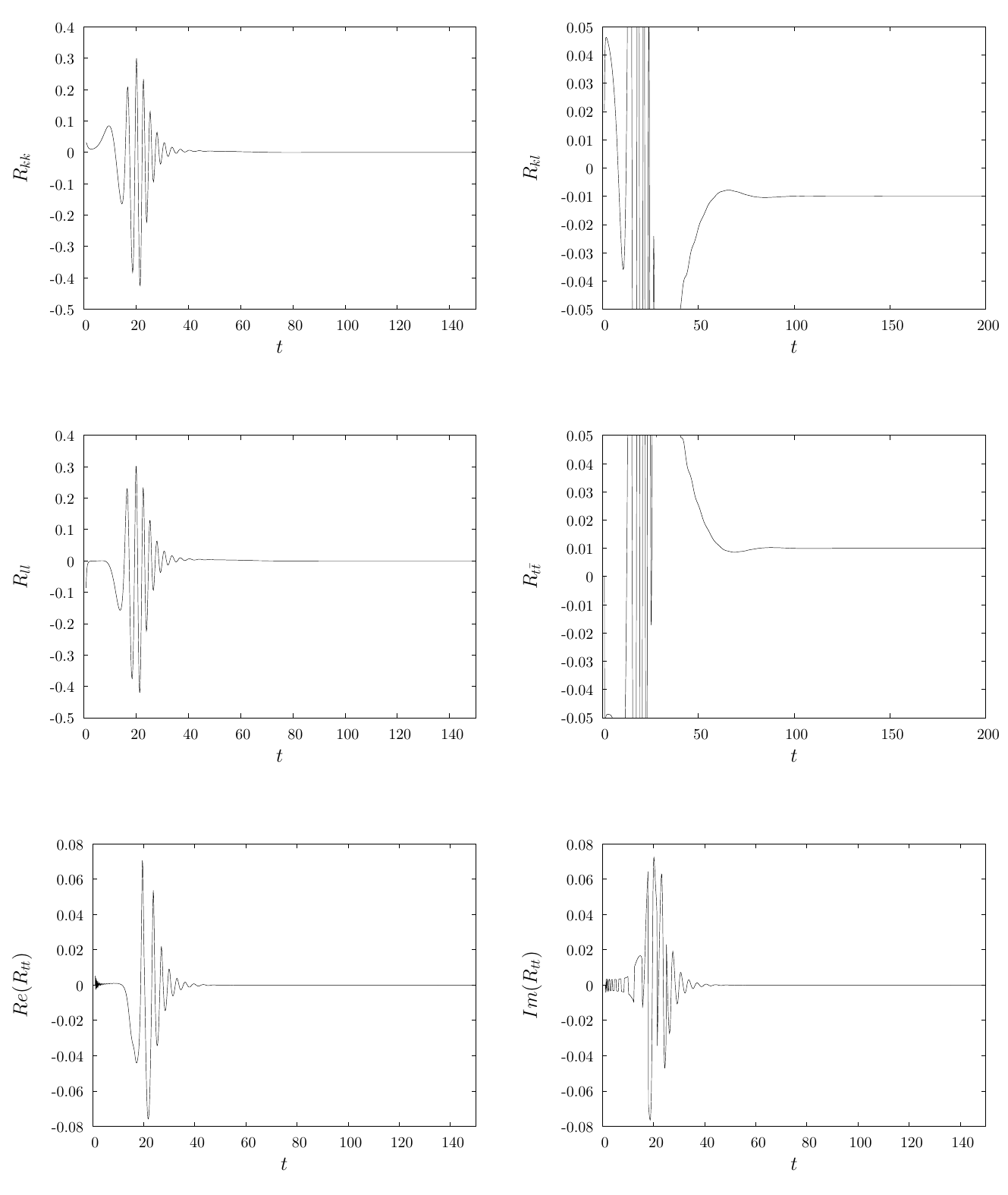}}
  \end{center}
  \caption{Numerical evolution of the non zero real and imaginary parts of the null tetrad \eqref{basenula} components of the Ricci tensor \eqref{riccinull}. The integration was done in the proper time. It can be seen that the Ricci tensor is asymptotically proportional to the null metric \eqref{metricanula}.}
  \label{ltricci}
\end{figure}

\begin{figure}[htpb]
    \begin{center}
  \resizebox{\imsize}{!}{\includegraphics{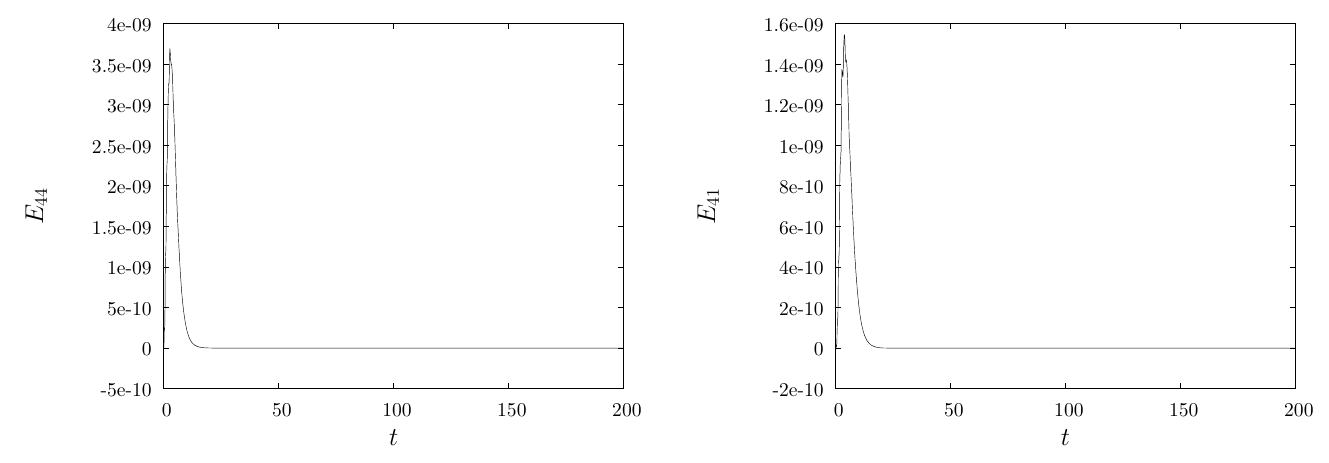}}
  \end{center}
  \caption{Numerical evolution of the constraints $E_{44}$ and $E_{41}$ which should be zero identically, indicating that the numerical solution is accurate. The integration was done for the initial condition and parameters specified on the text and until the proper time $200$.  }
  \label{figura2l}
\end{figure}

The time evolution of the non zero components of the Ricci tensor according to the null base \eqref{basenula} are shown in Fig.  \ref{ltricci}. For the particular metric chosen, \eqref{eq homogeneous and anisotropic line element} -\eqref{eq line element for spatially homogeneous four-space again}, $R_{kt}=R_{lt}=0$. 
It is also shown in Fig.  \ref{ltricci} that asymptotically, the Ricci tensor is a constant proportional to the null metric \eqref{metricanula}. 
The only consistent result with the field equation \eqref{eq field equation for semiclassical theory} is 
\[  
R_{ab}=\frac{\Lambda}{2}\left(\begin{array}{cccc}
0&-1 & 0 & 0\\
-1& 0 & 0&0\\
0 & 0&0&1\\
0&0 &1&0
\end{array}\right).
\]
We have numerically checked up to $t=500$ in proper time that $R_{ab}=\Lambda/2 g_{ab}$ with one part in $10^8$. Also the constraints $E_{44}=0$ and $E_{41}=0$ where numerically verified in Figure \ref{figura2l} which is a strong indication that the numerical result should be trusted.

\subsection{Asymptotically Riemann Flat Solution}

In the numerical solutions the following values for the parameters $C_1=0.3$, $C_2=2.0$, $A=0.1$, $\alpha=1000.0$, $\beta=-5.0$, $\Lambda=0$, where chosen, and initial conditions ($t=0$) near Minkowski solution \eqref{eq Juliano exact solution} 
\begin{eqnarray}
&&a_1(t)=At+C_1=C_1\nonumber\\
&&a_2(t)=C_2(At+C_1) -0.0015=C_2C_1-0.0015\nonumber\\
&&a_3(t)=C_2(At+C_1)+4.0\times 10^{-4}=C_2C_1+4\times 10^{-4} \nonumber\\
&&a_4(t)=0.0,
\label{condini}
\end{eqnarray}
we stress that numerically the initial condition $a_4=0$ but it rapidly increases. The metric does not remains diagonal. 

The only non null initial conditions for the derivatives are given in following Table \ref{tabela2}, and as before, the numerical values of  $d^3a_1/dt^3$ and $d^3a_3/dt^3$ are consistent with the two constraints $E_{44}\equiv 0$ and $E_{41}\equiv 0$. 
 \begin{table}
 \caption{The only non null initial conditions together with \eqref{condini}, consistent with the $E_{44}$ and $E_{41}$ constraints \label{tabela2}}
\begin{tabular}{ccccc}
\hline
$\dot{a}_1(0)$ & $\dot{a}_2(0)$  & $\dot{a}_3(0)$ & $\dddot{a}_1$ & $\dddot{a}_3$\\ 
\hline
$A$ & $C_2A$  & $C_2A$  & $2.45037249\times 10^{-2}$ & $9.82173813\times 10^{-2}$,\\ 
\hline
\end{tabular} 
\end{table}

\begin{figure}[htpb]
  \begin{center}
   \resizebox{\imsize}{!}{\includegraphics{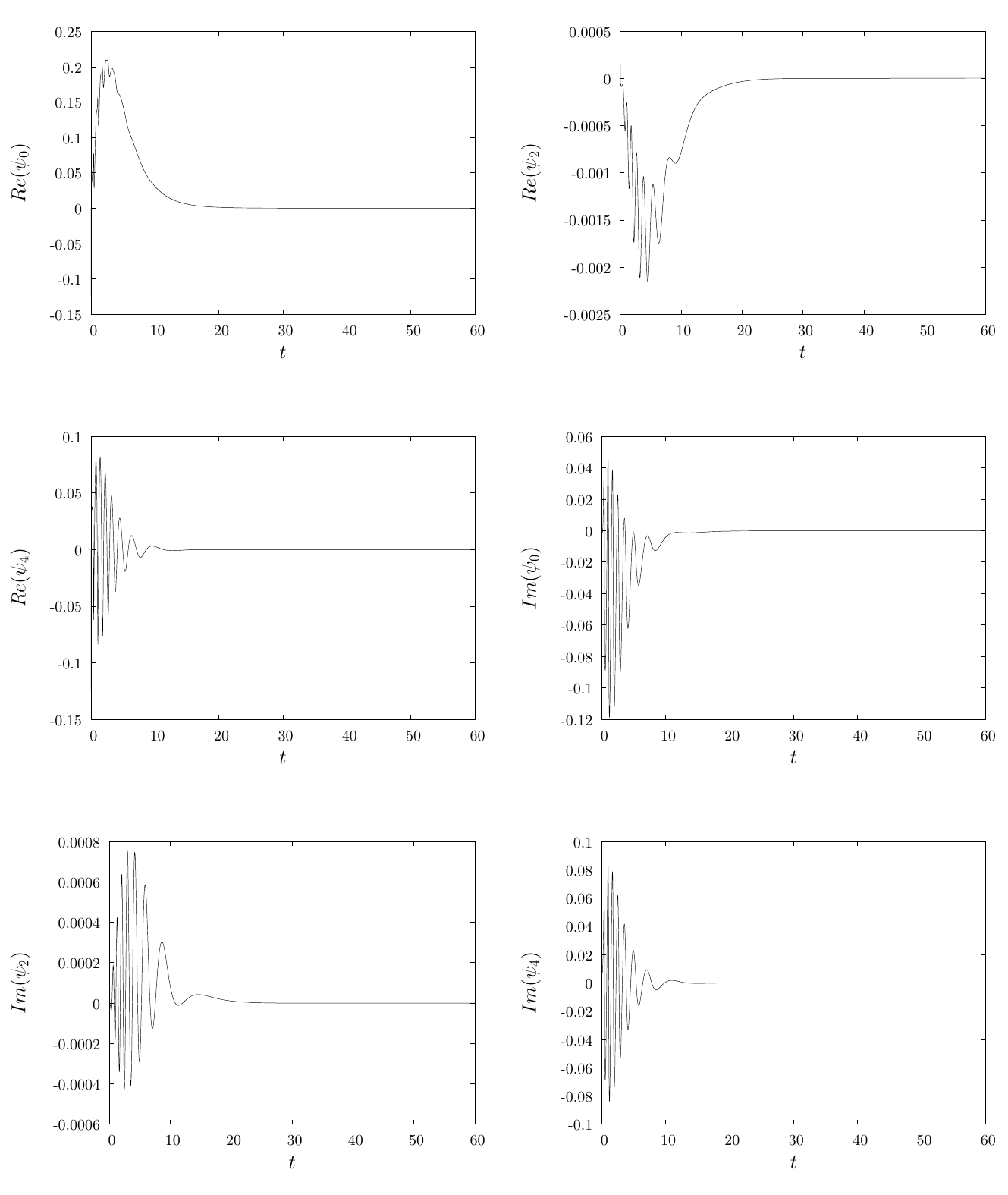}}
  \end{center}
  \caption{Numerical evolution of the real and imaginary parts of the Newman-Penrose coefficients $\psi_0$ - $\psi_4.$ For the initial condition given in the text. The numerical integration was done until the proper time $t=60$.}
  \label{figura1}
\end{figure}

\begin{figure}[htpb]
  \begin{center}
   \resizebox{\imsize}{!}{\includegraphics{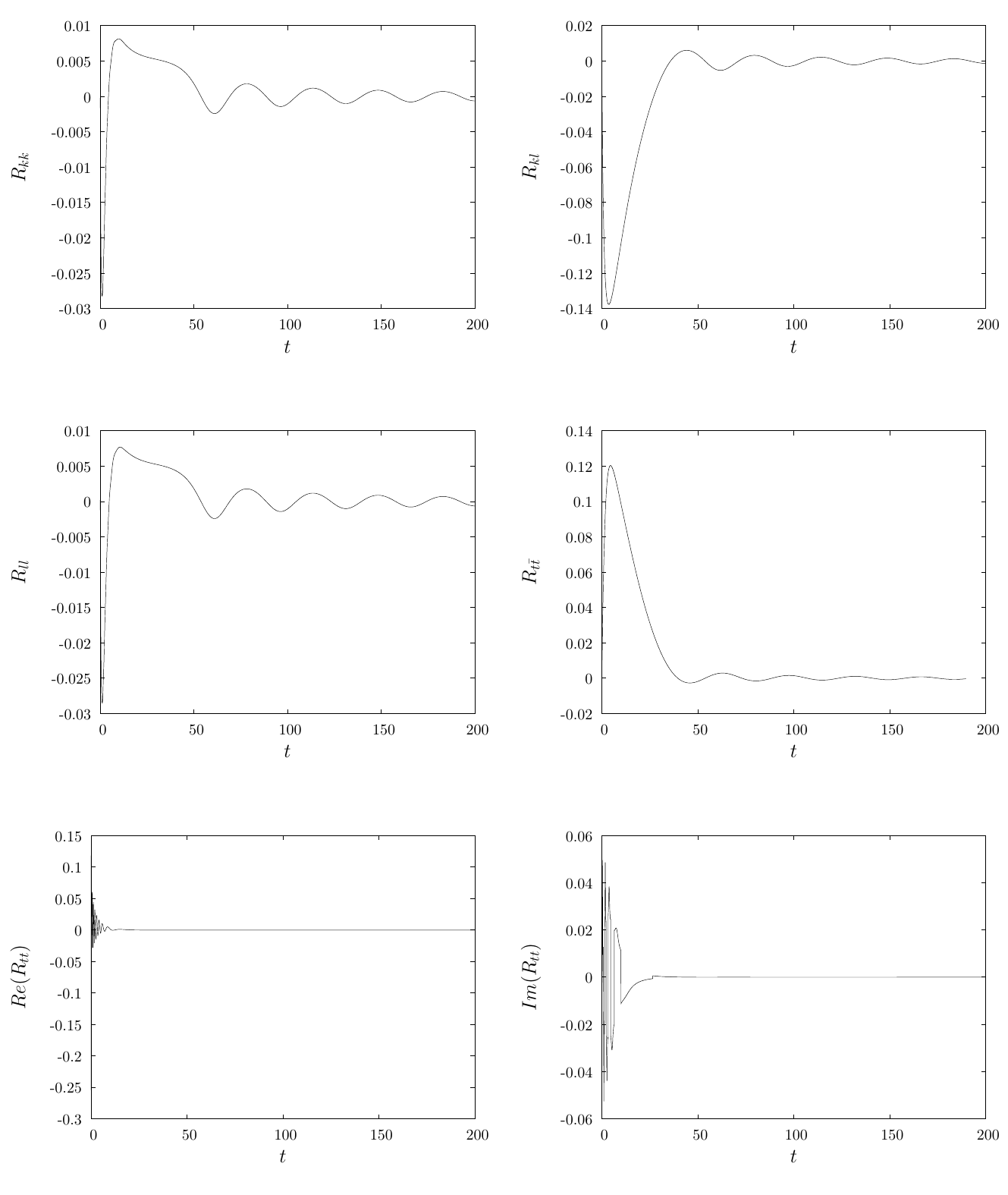}}
  \end{center}
  \caption{Numerical evolution of the non zero real and imaginary parts of the null tetrad \eqref{basenula} components of the Ricci tensor \eqref{riccinull}. The numerical integration was done until the proper time $t=200$.}
  \label{tricci}
\end{figure}

The time evolution of the non zero Newman-Penrose coefficients, $\psi_0$-$\psi_4$ are shown in Fig.  \ref{figura1}. For the line element chosen $\psi_1\equiv \psi_3\equiv 0$.
The time evolution of the non zero components of the Ricci tensor according to the null base \eqref{basenula} are shown in Fig.  \ref{tricci}. For the particular metric chosen, \eqref{eq homogeneous and anisotropic line element} -\eqref{eq line element for spatially homogeneous four-space again}, $R_{kt}=R_{lt}=0$. 
\newpage

\begin{figure}[htpb]
    \begin{center}
     \resizebox{\imsize}{!}{\includegraphics{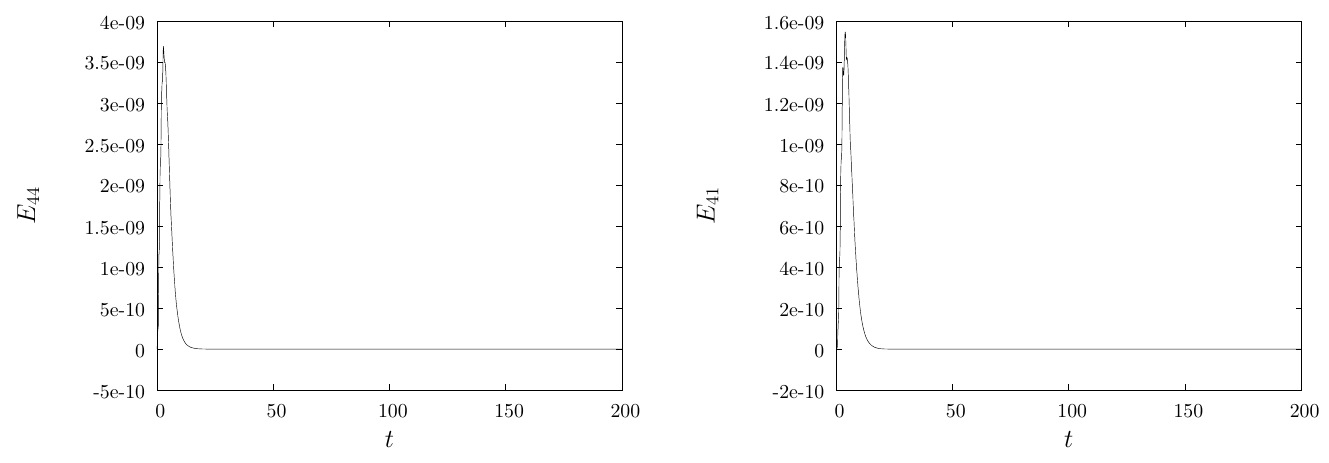}}

  \end{center}
  \caption{Numerical evolution of the constraints $E_{44}$ and $E_{41}$ which should be zero identically, indicating that the numerical solution is accurate. The integration was done for the initial condition and parameters specified on the text and until the proper time $t=200$.  }
  \label{figura2}
\end{figure}

The time evolution of the constrains $E_{44}$ and $E_{41}$ are shown in Fig.  \ref{figura2}.  As the constraints are numerically verified, we believe the numerical solution is accurate.

Since all the $\psi\rightarrow 0$, Fig.  \ref{figura1}, the Weyl tensor $C^a_{bcd}\rightarrow 0$, and since the Ricci tensor $R_{ab}\rightarrow 0$,   Fig.  \ref{tricci}, then the Riemann tensor $R_{abcd}\rightarrow 0$.
    
The numerical integrations shown in Fig.  \ref{figura1} and Fig.  \ref{tricci} were carried up to times $t=1\times 10^6$ showing that null tetrad components of the Riemann tensor $R_{abcd}\rightarrow 0$ in one part in $1\times 10^{6}$ and smaller values. It is in this sense that the solution is understood to asymptotically approach Minkowski space. 
\newpage
\subsection{Singularity}
Using exactly the same parameters of the preceding section  $C_1=0.3$, $C_2=2.0$, $A=0.1$, $\alpha=1000.0$, $\beta=-5.0$, $\Lambda=0$, a slightly different initial condition is chosen 
\begin{eqnarray}
&&a_1(t)=At+C_1=C_1\nonumber\\
&&a_2(t)=C_2(At+C_1) +0.2=C_2C_1+0.2\nonumber\\
&&a_3(t)=C_2(At+C_1)+4.0\times 10^{-4}=C_2C_1+4\times 10^{-4}\nonumber\\
&&a_4(t)=0.0,
\label{condsing}
\end{eqnarray}
we stress that numerically the initial condition $a_4=0$ but it rapidly increases. The metric does not remains diagonal. The only non null initial conditions for the derivatives are given in following Table \ref{tabela3}. 
 \begin{table}
 \caption{The only non null initial conditions together with \eqref{condsing}, consistent with the $E_{44}$ and $E_{41}$ constraints \label{tabela3}}
\begin{tabular}{ccccc}
\hline
$\dot{a}_1(0)$ & $\dot{a}_2(0)$  & $\dot{a}_3(0)$ & $\dddot{a}_1$ & $\dddot{a}_3$\\ 
\hline
$A$ & $C_2A$  & $C_2A$  & $-5.03312185\times 10^2$ & $-2.01311859\times 10^3$,\\ 
\hline
\end{tabular} 
\end{table}

\begin{figure}[htpb]
  \begin{center}
   \resizebox{\imsize}{!}{\includegraphics{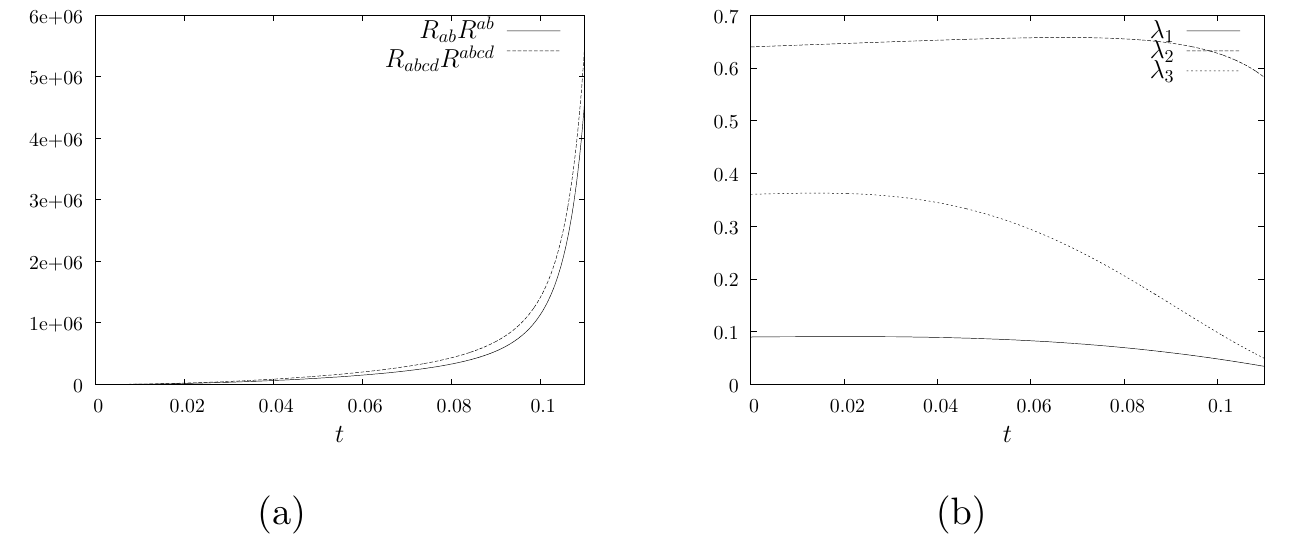}}
    \end{center}
    \caption{a) The increase of the scalar curvature invariants $R_{ab}R^{ab}$, $R_{abcd}R^{abcd}$ characterizing a singularity. b) The principle values of the metric showing a decrease of all the spatial directions, which we understand as a collapse. Where $\lambda_1=g_{11}=a_1^2$, $\lambda_2=[a_2^2+a_3^2+\sqrt{\Delta}]/2$,  $\lambda_3=[a_2^2+a_3^2-\sqrt{\Delta}]/2$ and 
 $\Delta=a_3^4-2a_2^2a_3^2+a_2^4+4a_4^2$. 
    \label{sing}}
    \end{figure}

\begin{figure}[htpb]
  \begin{center}
   \resizebox{\imsize}{!}{\includegraphics{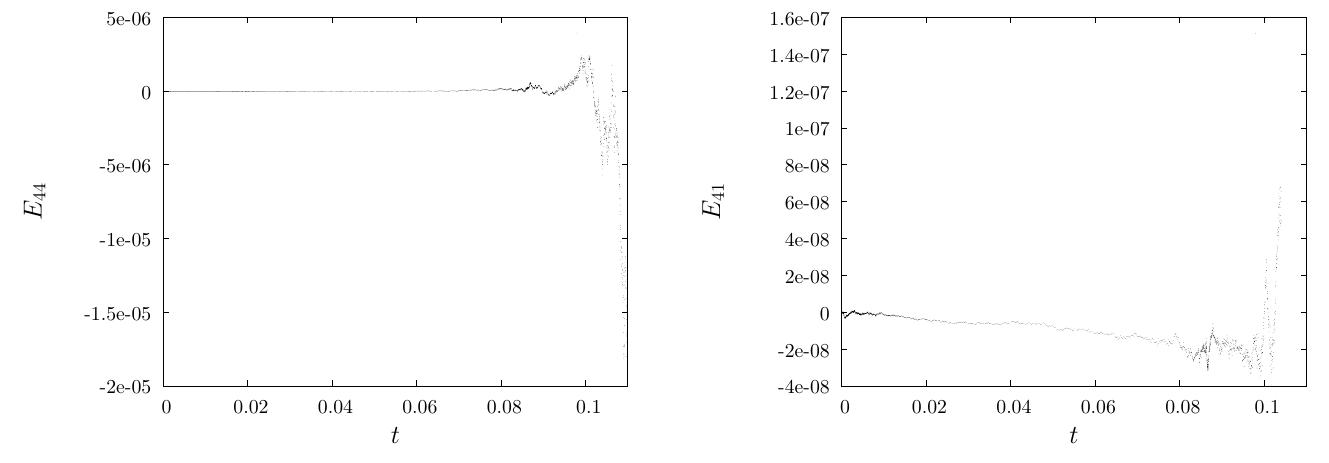}}
    \end{center}
    \caption{The constraints $E_{44}$ and $E_{41}$. Note that as the singularity is approached, the numerical errors increase, which is expectable. 
    \label{controle}}
    \end{figure}

This initial condition evolves very fast to a singularity characterized by the increase of the curvature scalars $R_{ab}R^{ab}$ and $R_{abcd}R^{abcd}$ shown in Fig.  \ref{sing}.
The constraints which should be zero $E_{44}$ and $E_{41}$ shown in Fig.  \ref{controle}, prove that the numerical result is accurate. Note that as the singularity is approached, the numerical errors increase, which is expectable. Also, the numerical integration can be carried further and further, and the constraints are not really satisfied at all, and is not shown since we believe the result should not be trusted. 
\newpage
\section{Conclusions}
In this present work, the numerical solutions of Bianchi $VII_A$ type are analyzed in the effective gravity context. This is the anisotropic generalization 
of the hyperbolic $H^3$ Friedmann ``open" space. The quadratic gravity is the result of vacuum polarization counter terms which must be introduced into Einstein's theory of gravitation. It should be the most natural theory just after the Planck era. It must be emphasized that although numeric the solutions are exact in the sense that they depend only on the precision of the machine. 

Within the cosmological context it is of great interest the concept of isotropisation, in that an initially irregular Universe turns out less irregular by physical processes. Or at least, the irregularities do not grow indefinitely. The supposition of a homogenous Universe is a very strong one, and allowing for cosmological anisotropy provides a less restricted initial condition than the isotropic Friedmann ``open" $H^3$ model. As a less restricted initial condition makes it a more probable one for the creation of the Universe, whatever this means. We can conclude that weak isotropisation occurs in the sense that the anisotropies either vanish or at least do not grow indefinitely, depending on the initial condition, of course. According to quadratic gravity, not all initial conditions evolve to reasonable Universes. There are initial conditions that evolve to a singularity. 

Apparently, the question of stability of de Sitter solution in quadratic gravity was first obtained by M\"uller, Schmidt and Starobinsky \cite{VMuller88}.

We have previously,  numerically investigated Bianchi $I$ type solutions \cite{sandro}. We found that there is a basin of attraction to Minkowski space for $\Lambda=0$ and a basin of attraction to de Sitter space for $\Lambda>0$, see \cite{daniel-sandro}. In this sense, Minkowski and de Sitter solutions are structurally stable according to the effective gravity, for the particular Bianchi $I$ models we analyzed. Also in the Bianchi $I$ case, there is a basin of attraction to a singularity. 

The numerical solutions are given in section 3. In section 3.1 the initial condition is chosen near the de Sitter exact solution. In section 3.2 and 3.3 the solution is chosen near Minkowski exact solution. Albeit initially the metric is supposed diagonal in all cases studied, it gets non diagonal dynamically through the $a_4(t)$ component, see eq. \eqref{eq homogeneous and anisotropic line element}.

The numerical solutions are characterized asymptotically according to the null tetrad components of the Weyl tensor and Ricci tensor. We emphasize that this is not such a strong criteria as the one discussed in the end of section 2. According to this criteria, for instance, Kasner's solution \cite{Kasner} asymptotes Minkowski space. 

In this sense, for the first solution under consideration in section 3.1, the Weyl tensor is zero and the Ricci tensor is proportional to the metric $R_{ab}=\Lambda/2g_{ab}$ asymptotically and we identify this solution with de Sitter space. In section 3.2 the solution asymptotes a Riemann flat space in sense that the Weyl tenosr and Ricci tensor vanish asymptotically. The convergence to Riemann flat space is not so fast and we shall not address this question in this present work. In section 3.3 a solution that converges to a singularity is presented.  

The existence of initial conditions that converge to a singularity, for instance the one given in section 3.3, is a strong indication that this quadratic gravity is certainly not a complete theory.

In all the cases we considered, the numerical behavior described in the text was checked for much larger times that the ones plotted, and we believe that asymptotical interpretation is the correct one. 

Since the initial conditions are near de Sitter and Minkowski solutions, we can speculate that also for Bianchi $VII_A$ solutions,  de Sitter and Minkowski space should be structurally stable in the weak sense specified near eq. \eqref{criterio}, according to the effective gravity, such that there should be basins of attraction to these solutions. We intend to address this stability issue in a future works. 

\begin{acknowledgements}
J. A. de Deus wishes to thank the Brazilian agency CNPq for
financial support. D. M. wishes to thank Brazilian projects: {\it
Nova F\'\i sica no Espa\c co} and INCT-A. The authors wish to thank an anonymous
referee for comments and suggestions.
\end{acknowledgements}

\end{document}